# MULTIPLE FRONT PROPAGATION IN A POTENTIAL NON-GRADIENT SYSTEM

M. SAN MIGUEL, R. MONTAGNE[1], A. AMENGUAL, E. HERNÁNDEZ-GARCÍA.
*Departament de Física*
*Universitat de les Illes Balears*
*E-07071 Palma de Mallorca, Spain*

ABSTRACT. A classification of dynamical systems in terms of their variational properties is reviewed. Within this classification, front propagation is discussed in a non-gradient relaxational potential flow. The model is motivated by transient pattern phenomena in nematics. A front propagating into an unstable homogenous state leaves behind an unstable periodic pattern, which decays via a second front and a second periodic state. An interface between unstable periodic states is shown to be a source of propagating fronts in opposite directions.

## 1. Introduction.

In this paper we will consider pattern formation in systems which approach, asymptotically in time, an homogeneous stable state. However, they exhibit long-lived pattern dynamics as transient states. Such patterns may originate in finite wavenumber fluctuations triggering the decay of an initially homogeneous stable state[1,2], or be a consequence of fronts propagating through the system[3]. We will focus here on patterns created by front propagation into unstable states. A physical motivation for this study is the long lived transient patterns observed in the Fréedericksz transition in nematic liquid crystals[4].

The type of systems discussed here are described by dynamical systems for which a Lyapunov functional, also called potential, exists. Time evolution proceeds then by minimizing such potential. However, we wish to emphasize that potential minimization can be attained along many different directions in the phase space of the system, depending on other ingredients of the dynamical system beyond the potential. In fact, the existence of rich transient dynamical behavior in many systems can be traced back to these other dynamical ingredients of the potential system. For pedagogical purposes we review in Sect. 2 a classification of dynamical systems in which the different role of relaxational vs. non-relaxational (non-dissipative) contributions to the dynamics is pointed out.

A general model for transient pattern dynamics is introduced in Sect. 3. The model is motivated by the physics of the Fréedericksz transition in nematics and justified from the nematodynamic equations[5]. Within the classification of Sect. 2, the model belongs to the class of potential and relaxational non-gradient flows. The non-gradient contributions originate in an approximation to non-relaxational dynamics associated with hydrodynamic coupling in the nematodynamic equations. In Sects. 4 and 5 we study front propagation in this model. A front connecting homogeneous stable and unstable states is seen to propagate leaving behind it a periodic spatial pattern which is linearly unstable. The decay of this pattern is via an intermediate periodic pattern of larger wavelength. This second pattern is still linearly unstable, but it is more stable in terms of the Lyapunov potential. It appears behind a second front which propagates into the original periodic state. The velocity of the two propagating fronts, as well as the wavenumber of the periodic states, is successfully determined from marginal stability arguments[6]. Finally, in Sect. 5 we consider the fronts which originate in the interface between two periodic unstable states of different periodicity. Such interface is shown to be a source of fronts propagating in opposite directions while the system evolves through states of greater stability in its search for the global potential minimum.

---
[1]on leave from Universidad de la República (Uruguay).

## 2. What is a potential system?

It is commonly stated that the rich variety of dynamical states that occur in non-equilibrium systems originates in the non-potential or non-variational character of the dynamical models which describe them. The main argument is that, in these cases, the study of the dynamics can not be reduced to the minimization of a potential which plays the role of the free energy of equilibrium systems. This general statement needs to be qualified, since it is well known that models used to study equilibrium critical dynamics[7] include mode-mode coupling terms such that its dynamical evolution is not simply given by the minimization of the free energy. In the following we review a classification of dynamical systems that, although well established in the context of stochastic dynamics[8,9] it is often overlooked in general discussions of deterministic spatio-temporal dynamics.

Non-potential dynamical systems are usually defined as those for which there is no Lyapunov potential giving the time evolution. Unfortunately, this definition is also applied to cases in which there is no *known* Lyapunov potential. To be more precise, let us consider dynamical systems of the form

$$\dot{\psi} = A[\psi] \tag{2.1}$$

where $\psi$ represents the set of dynamical variables. For generality, we take them to be complex, and the notation $\psi^*$ represents the complex conjugate of $\psi$. Here the dynamical variables will be spatially dependent fields: $\psi = \psi(\mathbf{x}, t)$. $A[\psi]$ is a functional of them. Let us now split $A$ into two contributions:

$$A[\psi] = G[\psi] + N[\psi], \tag{2.2}$$

where $G$, the *relaxational* part, will have the form

$$G[\psi] = -\Gamma \frac{\delta F[\psi]}{\delta \psi^*}, \tag{2.3}$$

with $F$ a real and scalar functional of $\psi$. $\Gamma$ is an arbitrary hermitic and positive-definite operator (possibly depending on $\psi$). In the case of real variables there is no need of taking the complex conjugate, and hermitic operators reduce to symmetric ones. The functional $N[\psi]$ in (2.2) is the remaining part of $A[\psi]$. The important point is that, if the splitting (2.2) can be done in such a way that the following orthogonality condition is satisfied (c.c. denotes the complex conjugate expression):

$$\int d\mathbf{x} \left( \frac{\delta F[\psi]}{\delta \psi^*} N[\psi(\mathbf{x})]^* + \text{c.c.} \right) = 0, \tag{2.4}$$

then the terms in $N$ neither increase nor decrease the value of $F$, which due to the terms in $G$ becomes a decreasing function of time:

$$\frac{dF[\psi(\mathbf{x},t)]}{dt} \leq 0 \ . \tag{2.5}$$

If $F$ is bounded from below then it is a Lyapunov potential for the dynamics (2.1). Equation (2.4), with $N = A - G$ can be interpreted as an equation for the potential $F$ associated to a given dynamical system (2.1). It has a Hamilton-Jacobi structure. Its solution is in general a difficult task, but several non-trivial results exist in the literature[8,10].

Once this notation has been set-up, we can call relaxational systems those such that there is a solution $F$ of (2.4) such that $N = 0$, that is all the terms in $A$ contribute to decrease $F$. Potential systems can be defined as those for which there is a nontrivial (i.e. a non-constant) solution $F$ to (2.4). A more detailed classification is the following:

1.- **Relaxational gradient flows:** Those dynamical systems for which $N = 0$ and $\Gamma$ is a constant. In this case the time evolution of the system follows the lines of steepest descent of $F$. A well known example is the Fisher-Kolmogorov equation, also known as model A of critical dynamics[7], or (real) Ginzburg-Landau equation for a real field $\psi(\mathbf{x}, t)$:

$$\dot{\psi} = r\nabla^2 \psi + c\psi - b \mid \psi \mid^2 \psi \ , \tag{2.6}$$

where $r$, $c$, and $b$ are real coefficients. This equation is of the form of Eq. (2.1)-(2.3) with $N = 0, \Gamma = 1$, and $F = F_{GL}[\psi]$, the Ginzburg-Landau free energy:

$$F_{GL}[\psi] = \int d\mathbf{x} \left( \frac{r}{2} \mid \nabla\psi \mid^2 - \frac{c}{2} \mid \psi \mid^2 + \frac{b}{4} \mid \psi \mid^4 \right) \tag{2.7}$$

2.- **Relaxational non-gradient flows:** Still $N = 0$, but $\Gamma$ is not constant, so that the dynamics does not follow the lines of steepest descent of $F$. A well known example of this type is the Cahn-Hilliard equation of spinodal decomposition, or model B of critical dynamics[7]:

$$\dot{\psi} = (-\nabla^2) \left( -\frac{\delta F_{GL}[\psi]}{\delta \psi} \right) \ , \tag{2.8}$$

The symmetric and positive-definite operator $(-\nabla^2)$ has its origin in a conservation law for $\psi$.

3.- **Non-relaxational potential flows:** $N$ does not vanish, but the potential $F$, solution of (2.4) exists and is non-trivial. Most models used in equilibrium critical dynamics[7] belong to this category. A simple example of this is

$$\dot{\psi} = -(1+i)\frac{\delta F_{GL}[\psi]}{\delta \psi^*} \ , \tag{2.9}$$

where now $\psi$ is a complex field. Notice that we can not interpret this equation as being of type 1, because $(1+i)$ is not a hermitic operator, but still $F_{GL}$ is a Lyapunov functional for the dynamics. Equation (2.9) is a special case of the Complex Ginzburg- Landau Equation

(CGLE), in which $A[\psi]$ is the sum of a relaxational gradient flow and a nonlinear Schrödinger-type equation $N[\psi] = -i\frac{\delta F_{GL}[\psi]}{\delta \psi^*}$. The general CGLE[11] is of the form (2.6) but $\psi$ is complex and $r$, $c$, and $b$ are arbitrary complex numbers. Calculations by Graham and coworkers indicate[12,13] that the CGLE, a paradigm of complex spatio-temporal dynamics, might be classified within this class of non-relaxational potential flows. The difficulty is that the explicit form of the potential to be obtained as a solution of (2.4), is only known in an uncontrolled small-gradient expansion.

4.- **Non-potential flows:** Those for which the only solutions $F$ of (2.4) are the trivial ones (that is $F$ = constant). Hamiltonian systems as for example the nonlinear Schrödinger equation are of this type.

In this paper we only deal with potential situations. The model we introduce in the next section belongs to the class of relaxational non-gradient dynamics. The non-gradient character has its origin in the non-relaxational and more complex dynamics of nematic liquid crystals in a magnetic field, from which our model is obtained after some approximations.

### 3. A general model for transient pattern dynamics

Transient pattern formation is well documented experimentally for different instabilities in nematic liquid crystals[14]. We discuss here a general model whose physical motivation is the magnetic Fréedericksz transition. In this transition the reorientation of the nematic director in response to a large enough applied magnetic field does not proceed homogeneously: a transient striped pattern with a characteristic wavelength emerges. At long times the pattern disappears leading to the homogeneously reoriented final equilibrium state. We consider a twist geometry in which the sample is contained between two plates separated a distance $d$ and perpendicular to the z-axis. The nematic material is prepared with its director field $\vec{n}^o$ aligned with the x-axis and a magnetic field $\vec{H}$ is applied in the y-direction. When the magnetic field is switched-on at time $t = 0$ from an initial value smaller than a critical value to a final value above it a striped pattern appears in the x-y plane with domain walls parallel to the y-axis.

The dynamics of this system can be described in terms of the nematodynamic equations. We assume homogeneity in the y-direction and that the reorientation takes place in the x-y plane, so that the director field $\vec{n}(\vec{r})$ is written in terms of an angle $\phi$ as $n_x(x, z) = \cos\phi(x, z)$, $n_y(x, z) = \sin\phi(x, z)$. The director field is coupled to a velocity field $\vec{v}(x, z)$ which in the geometry described above we assume to be oriented along the y-direction. In a minimal coupling approximation the equations for $\phi$ and $v_y$ become[15]:

$$d_t\phi(x, z) = -\frac{1}{\gamma_1}\frac{\delta F}{\delta \phi} + \frac{1}{2\rho}(1 + \lambda)\partial_x \frac{\delta F}{\delta v_y} , \tag{3.1}$$

$$d_t v_y(x, z) = \frac{1}{2\rho}(1 + \lambda)\partial_x \frac{\delta F}{\delta \phi} + \frac{1}{\rho^2}(\nu_2 \partial_z^2 + \nu_3 \partial_x^2)\frac{\delta F}{\delta v_y} . \tag{3.2}$$

$\rho$ is the mean density, $\gamma_1$, $\lambda$, $\nu_2$, and $\nu_3$ are viscosity coefficients and $F$ is a free energy:

$$F = \int d\vec{r}\{\frac{1}{2}\left[k_1(\vec{\nabla} \cdot \vec{n})^2 + k_2(\vec{n} \cdot \vec{\nabla} \times \vec{n})^2 + k_3(\vec{n} \times (\vec{\nabla} \times \vec{n}))^2\right] - \frac{1}{2}\chi_a(\vec{n} \cdot \vec{H})^2 + \frac{1}{2}\rho\vec{v}^2\} \tag{3.3}$$

The first three terms in (3.3) form the Oseen-Frank free energy associated with distortion of the director field, with $k_1, k_2$ and $k_3$ being splay, twist and bend elastic constants respectively. The next term is the magnetic contribution with $\chi_a$ being the anisotropic susceptibility, and the last term gives the hydrodynamic contribution.

This dynamical model falls within the general category of non-relaxational potential flows discussed above. The free energy $F$ is a Lyapunov functional, but the dynamical model contains non-relaxational terms which give a vanishing contribution to the time derivative of $F$. The term proportional to $1/\gamma_1$ in (3.1) gives a gradient relaxational dynamics for $\phi$, and the terms proportional to $\nu_2$ and $\nu_3$ in (3.2) give a non-gradient relaxational dynamics associated with viscosity for the velocity flow. On the other hand, the terms proportional to $(1 + \lambda)$ in (3.1) and (3.2), which give the coupling between $\phi$ and $v_y$, are of non-dissipative hydrodynamical origin and produce a non-relaxational dynamics for the whole system.

Dynamics can be described in terms of the amplitude $\psi(x)$ of the most unstable Fourier mode of $\phi(x,z)$ in the z-direction. A useful approximation in the limit of small inertia is to eliminate adiabatically the velocity field, which leads to a closed equation for $\psi(x,t)$. In appropriate dimensionless units such equation reads[16]

$$\partial_t \psi(x,t) = \Gamma(\partial_x)[\partial_x^2 \psi + c\psi - b\psi^3], \ b,c > 0 \quad (3.4)$$

where $\Gamma(\partial_x)$ is a complicated kinetic coefficient which contains the remanent effect of the hydrodynamic coupling of director and velocity field after the adiabatic elimination of the latter. In the limit of long wavelengths that we consider in the following $\Gamma(\partial_x)$ becomes

$$\Gamma(\partial_x) \approx a - \partial_x^2, \ a > 0 \quad (3.5)$$

In the absence of hydrodynamic coupling $\Gamma$ becomes a constant and (3.4) describes a gradient flow. In general (3.4) describes a non-gradient relaxational dynamics, with the $\partial_x^2$ term in (3.5) being the leading contribution from the non-relaxational dynamics in (3.1)-(3.2). By comparison with (2.8) we call (3.4) Modified Cahn-Hilliard Equation (MCH).

Eqs. (3.4) and (3.5) define a generic model to study transient pattern dynamics[1]. For this study it is first important to recall the existence of at least three kinds of bounded stationary solutions: (a) $\psi(x) = \pm\sqrt{c/b} \equiv \psi_\pm$, (b) $\psi(x) = 0$, and (c) a family of periodic solutions $\psi_q(x)$ of fundamental wavenumber $q$ in the range $(0, \sqrt{c})$. Regarding stability with respect to small perturbations, both solutions included in type (a) are linearly stable, and all the solutions in (b) and (c) and linearly unstable. The dynamical evolution from the unstable solution (b) to the stable solution (a) proceeds via the formation of long-lived transient patterns which locally are close to solutions of type (c). The initial stages of pattern formation can be understood by the fact that the linear stability analysis of (3.4)-(3.5) around the solution $\psi = 0$ identifies a mode with finite wavenumber as the most unstable one. The instability becomes of zero wavenumber in the limit of gradient-flow dynamics in which $\Gamma$ is a constant. What we show in the next section is that the periodic solutions (c) can also be realized by front propagation: A front connecting solutions (a) and (b) advances into (b) and leaves behind it a solution of type (c). This solution is shown to decay via a secondary front which separates states close to two different solutions of type (c).

## 4. Fronts propagating into periodic unstable states

The description of a uniform stable state advancing into a uniform unstable one is well known[17,18]. It is also known that front propagation can produce unstable periodic patterns, as explicitly demonstrated for the Extended Fisher-Kolmogorov model (EFK)[19], which is a relaxational gradient system. For the MCH model a new situation occurs, as shown in Fig. 1.

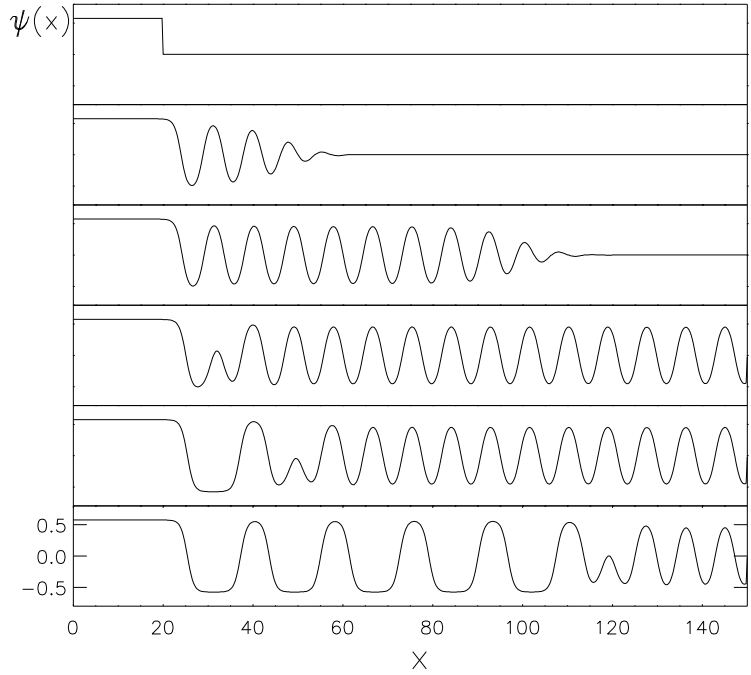

Figure 1 Graph of $\psi(x,t)$ for $t = 0, 20, 50, 125, 195$, and $455$ for $a = 0.2$ and $b = 3$.

First, a front advances into the initial homogenous state leaving behind a periodic state. The novelty is that a second front appears and moves into the periodic state leaving behind a second periodic state with different periodicity. This was observed for different values of $a$. Figure 2 shows space-time plots of $\psi(x,t)$ obtained from a numerical solution of eqs. (3.4-3.5) for several values of $a$. The horizontal axis represents space, and time runs along the vertical axis. White corresponds to regions with $\psi$ near the value of the stationary stable solution $1/\sqrt{b}$ and black to those with a value near $-1/\sqrt{b}$. The location of the fronts is clear and the measured slopes determine their velocities.

Our numerical results indicate that the mechanism of decay of the periodic state left behind the first front is similar to the decay of the initial $\psi = 0$ unstable state. That is, a front replaces the unstable state by some more stable state. Relative stability is here defined in terms of the Lyapunov functional of the model. This interpretation leads naturally to the study of fronts propagating into periodic unstable states as relevant for the understanding of the *second-front* phenomenon.

The first front, the one moving into the homogeneous phase, has a velocity well reproduced by marginal stability theory[19–21,6,18]. We will now show that the second front, understood as a front moving into an unstable periodic state, can also be described by a generalization of such theory[3]. There are several ways of formulating the marginal stability hypothesis for propagation into an homogeneous state. In all of them the dynamics of the front is analyzed in the leading edge, where the field is small enough so that the equation describing its evolution can be linearized. We use here the steepest descent (or saddle point) approach[19,22], since it is easily generalized to front propagation into a periodic state.

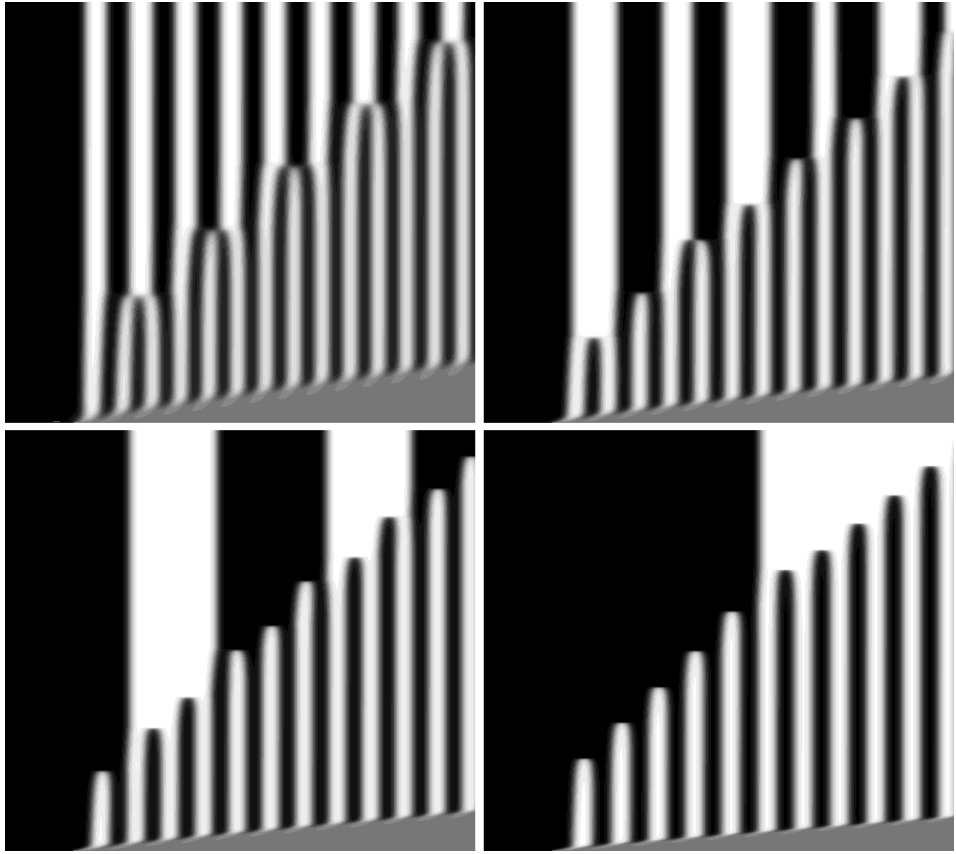

Figure 2 Space-time plots of $\psi(x, t)$ in gray levels, for $b = 3$ and $a = 0.0, 0.3, 0.6$ and $0.9$. Space is represented along the horizontal axis (system size: 137.5) and time runs along the vertical one from 0 to 575.

We assume that the periodic pattern left by the first front is close to one of the stationary solutions with dominant wavenumber $q = q_i$, that will be denoted by $\psi_{q_i}$. We linearize around the periodic state $\psi(x, t) \equiv \psi_{q_i}(x) + \delta\psi(x, t)$ obtaining an equation for $\delta\psi$ of the form

$$\delta\dot\psi = [\mathcal{L} + \mathcal{U}_{q_p}] \, \delta\psi \ . \tag{4.1}$$

$\mathcal{L}$ is the linear operator giving the dispersion relation corresponding to linearization around the uniform solution $\psi = 0$, and $\mathcal{U}_{q_p}$ stands for the remaining part: a periodic operator of periodicity $q_p$ related to $q_i$. Eq. (4.1) is a linear equation with periodic coefficients whose formal solution is given by Bloch (or Floquet) theory in terms of the eigenfunctions and eigenvalues of the linear operator

$\mathcal{L} + \mathcal{U}_{q_p}$. Given $\delta\psi_0(x) = \delta\psi(x, t = 0)$, the solution of (4.1) can be expressed in terms of the eigenfunctions $f_k(x)$ (of the Bloch form) and the eigenvalues $\epsilon(k)$ of the linear operator $\mathcal{L} + \mathcal{U}_{q_p}$

$$\delta\psi(x,t) = \int_{-\infty}^{+\infty} e^{\epsilon(k)t} f_k(x) \delta\hat{\psi}_0(k) dk \qquad (4.2)$$

where

$$\delta\hat{\psi}_0(k) = \int_L f_k^*(x) \delta\psi_0(x) dx \,, \qquad (4.3)$$

The integral is evaluated using the saddle point method in a referential frame $z = x - vt$ moving with the front velocity $v$,

$$\delta\psi(z, t \to \infty) \sim e^{h(q'_s)t + iq'_s z} \qquad (4.4)$$

where $h(q') = iq'v + \epsilon(k = q' - mq_p)$, and $q'_s$ is the dominant saddle point of the function $h(q')$ extended to the complex plane. In our case[3], $mq_p \equiv -2q_i$, so that $k = q' + 2q_i$. The saddle point condition and the additional requirement that the perturbation $\delta\psi$ remains finite and bounded in time in the vicinity of the leading edge ($z \sim 0$) lead to

$$\begin{cases} \left.\frac{d\epsilon(k)}{dk}\right|_{k=q'_s - mq_p} = -iv \\ \\ \text{Re}[h(q'_s)] = 0 \end{cases} \qquad (4.5)$$

From equation (4.5) with one can determine the velocity of the front $v = v_s$ and the complex number $q' = q'_s$ locating the saddle point. These equations are equivalent to the ones obtained for a front moving into an homogeneous state. The interpretation of $q'_s$ is the same as in such case: Eq.(4.4) shows that the real part of $q'_s$ gives the periodicity of $\delta\psi$ at the edge of the front, and its imaginary part characterize its steepness. The difference with the homogeneous case lies in the different eigenvalue spectrum $\epsilon(k)$.

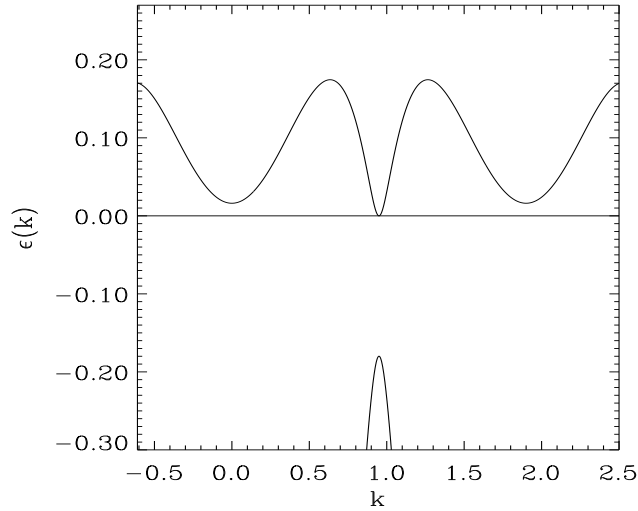

Figure 3 The spectrum $\epsilon(k)$ in the weak coupling approximation with $a = 0.02$, $c = 1.0$, $b = 3.0$ and an initial wavenumber $q_i = 0.95$

Figure 3 shows two branches of the spectrum $\epsilon(k)$ calculated using a *weak coupling approximation*[3] for an initial wavenumber $q_i = 0.95$. Since the upper branch is the positive one, it is the only to be used in (4.5).

The wavelength $\lambda$ of the periodic pattern left behind by the moving front can be calculated following a standard prescription[19]: we assume that the oscillations created at the leading edge by the linear instability will become quenched by the nonlinearities, but their periodicity will not be modified. In the moving frame of speed $v_s$, linear theory predicts that the leading edge (4.4) oscillates at a frequency such that a number of nodes $\Phi$ is created in the unit of time, with $\Phi = \pi^{-1}\left(\mathrm{Im}\left[\epsilon(k_s)\right] + v_s \mathrm{Re}\left[q'_s\right]\right)$. Behind the front, where the pattern has a wavelength $\lambda$, the flux of nodes passing in the unit of time through a point fixed in the moving frame is $2v_s/\lambda$. From this we get

$$\lambda^{-1} = \frac{1}{2\pi}\left(\frac{\mathrm{Im}\left[\epsilon(k_s)\right]}{v_s} + \mathrm{Re}\left[q'_s\right]\right) \; , \tag{4.6}$$

$\lambda$ is determined by the two different wavenumbers $k_s$ and $q'_s$. The hypothesis behind this formula are that no nodes are created nor destroyed far from the leading edge, and that every node that linear analysis predicts to be created has to be really created.

The velocities of the front propagating into the periodic state obtained from the theory and from numerical solution of the MCH equation are shown in Fig.4a. The mean periodicities left behind the second front are shown in Fig.4b together with those obtained from Eq. (4.6).

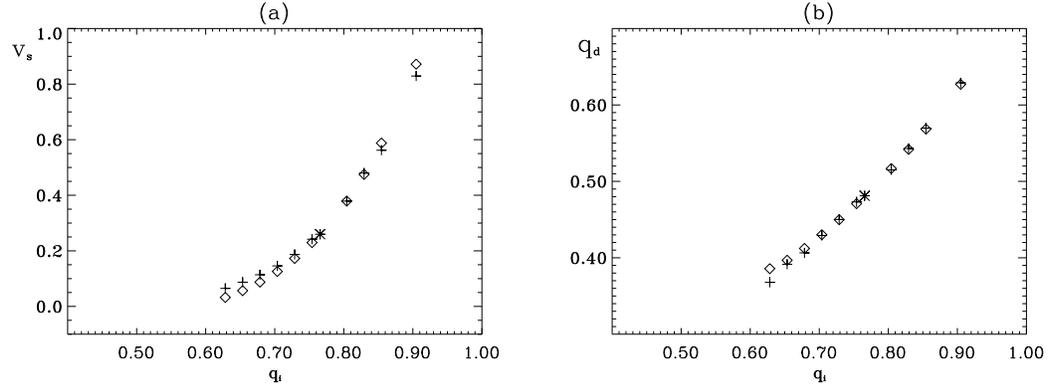

Figure 4 Velocity of the second front and wavenumber of the structure created when the front propagates into stationary states $\psi_{q_i}(x)$ with different dominant wavenumber ($a = 0$, $c = 1$). +: numerical simulations, $\diamond$ : theory. The point represented by $*$ corresponds to the value of $q_i$ obtained in the wake of the first front.

The initial conditions $\psi_0(x)$ for the numerical simulations were states in which a part of the system is in the $\psi_+$ state and the rest in the periodic unstable state $\psi_{q_i}(x)$ state. Simulations with several values of $q_i$ are shown in Fig. 4. The $\psi_{q_i}(x)$ states, which are periodic functions containing the mode $q_i$ and its harmonics, were obtained numerically by integrating the MCH equation with an approximation to the stationary solution as initial condition: $\psi_{q_i}(x) \approx [4(c - q_i^2)/3b]^{1/2}\sin(q_i x)$.

For $q_i > 0.9$ the pattern is already very unstable and numerically it decays by roll annihilation before the appearance of the front. On the other hand, for $q_i < 0.6$ the velocity of the front tends to be so small that the computer time needed to observe it becomes prohibitive. For intermediate $q_i$,

Fig. 4 shows good agreement between theoretical and numerical values for the velocity of the front and the wavenumber of the periodic pattern left behind (determined from the average wavelength of the pattern). Specially for $q_i \lesssim \sqrt{c}$, for which the weak coupling approximation is justified.

Once the theory for front propagation into periodic unstable states has been proved to be accurate, we used it to describe the second-front phenomenon of Figs. 1 and 2, that is, to predict the speed and the periodicity left behind a front that advances into the periodic state created by a first front, which is slightly different from the $\psi_{q_i}(x)$ used before. For small $a$ the agreement is good[3] and becomes poorer for large $a$. The reason is that for large $a$ the pattern created by the first front has a wavenumber quite different from $\sqrt{c}$, so that the weak coupling approximation used for calculating $\epsilon(k)$ is not accurate. It is interesting to note that for nematics such as PBG in solution[23], and for applied magnetic fields of about 8 kG, the parameters in the MCH model are $b \approx 3$, $c \approx 1$, and $a \approx 0.02$, which are in the range of validity of the theory. The theoretical prediction in such case is that the speed and periodicity behind the first front are of the order of 6 $\mu$m/s and 100 $\mu$m, respectively, whereas for the second front the values predicted are approximately 1$\mu$m/s and 170$\mu$m. Experiments to check these predictions would be welcome.

It is finally interesting to point out analogies and differences between the MCH model (a relaxational non-gradient flow) studied here and the EFK model[20] (a relaxational gradient flow). In the decay of an initial homegeneous unstable state a transient pattern with selected periodicity occurs in the MCH model, while the zero wavenumber is the most unstable mode in the EFK model. In front propagation into the unstable homogeneous state a pattern behind the front appears in both models. However, the second front described here possibly occurs in the EFK model only in situattions which are very difficult to observe numerically.

## 5. Interface between periodic unstable patterns: A source of propagating fronts

The results obtained in the previous section can be understood by saying that within the non-gradient potential flow (3.4)-(3.5), an unstable periodic state in contact with the homogeneously stable state decays through intermediate periodic and linearly unstable states. Such intermediate states are more stable than the initial one in terms of the Lyapunov potential of the problem. In addition the validity of the marginal stability criterion implies that the velocity of the front only depends on the initial periodic state and not on the state originally on the stable side of the front. To check the generality of these ideas we have considered the evolution of an interface between two periodic states of different wavenumbers $q_1 < q_2$. Both are linearly unstable, but from the argument above we expect the generation of a front moving into the less stable state (the one with larger wavenumber $q_2$) and leaving behind it a third state of wavenumber $q_3 < q_2$. If this process is still described by the marginal stability criterion the front velocity should be independent of $q_1$.

Numerical results corroborating such expectations are shown in Fig. 5. We note that the interface is far away from the boundaries of the system and, as the propagation of the front is analyzed also far enough from them, the boundaries of the system are meaningless in this discussion. We also note that two periodic solutions can be joined either through their maxima or through points of zero amplitude. In the first case, the initial condition will have a small jump since the steady amplitude depends on $q_i$. In the second case the change is smoother. However, when the matching is done with the homogeneous steady solution only the first case is meaningful. From Fig. 5 it is apparent that the way in which the matching is done does not affect the steady movement of the front: despite there is a slightly different delay for the first white region to disappear, the velocity of both fronts is the same and equal to 0.22. To check the independence of front velocity on $q_1$ we have also considered the situations with $q_1 = 0.785, 0.628$, and $0.419$ joined to a region with $q_2 = 0.698$. In all cases we have found the same front velocity ($0.21 \pm 0.01$) propagating into the region of $q_2$.

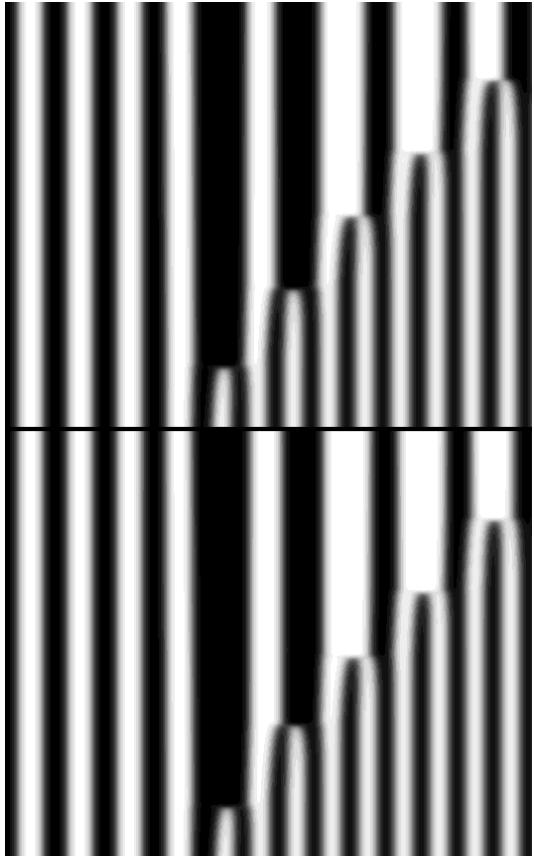

Figure 5 Space-time plots of $\psi(x,t)$ in gray levels (system size=130, the time for both plots run from 0 to 500). The initial condition correspond to an interface between two periodic stationary solutions with $q_1 = 0.503$ and $q_2 = 0.698$ connected through their maxima (top) and through a zero amplitude point (bottom)

The iteration of the mechanism that we have described of front propagation between two unstable periodic states naturally leads to a source of fronts propagating in opposite directions: if in the example in Fig. 5 it turns out that $q_1 > q_3$ we should expect a new front, now moving into the $q_2$-region. Such behavior of an interface as a source of fronts is seen in Fig. 6. We have considered here $q_1 = 0.698 < q_2 = 0.785$. A first front appears moving to the right with a velocity of 0.43. The pattern left behind by this front has a dominant wavenumber $q_3 < q_1 = 0.698$. The situation is then similar to the one found in Fig. 5 where we had a front propagating into a region with wavenumber 0.698, and, indeed, we find a front moving towards the left with the expected velocity of 0.21. This front leaves behind a new state with dominant wavenumber $q_4$ and a new front should emerge in the interface between the $q_2$ and $q_4$ regions already created by front propagation.

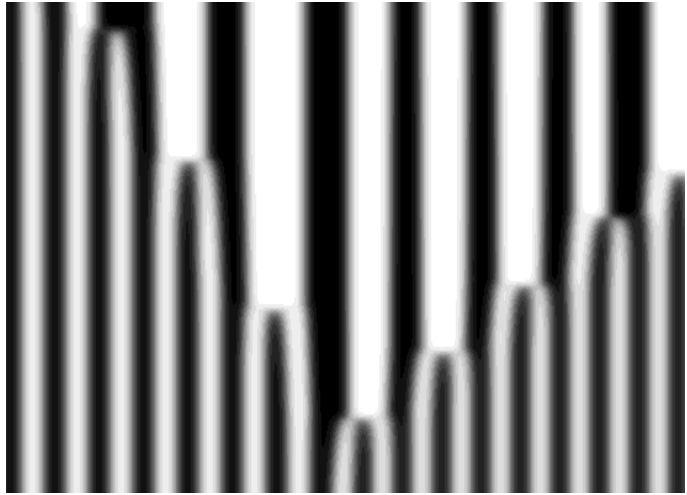

Figure 6 Space-time plots of $\psi(x,t)$ in gray levels. Two fronts moving in opposite directions emerge at the interface between the periodic patterns ($q_1 = 0.698$ and $q_2 = 0.785$; (system size=137.5 and time from 0 to 300).

We finally note that, in practice, the patterns left behind a front are not perfectly periodic. This makes the decay of periodic unstable states by a bulk mechanism more efficient and limits the number of successive fronts that are observed. In any case, and as a speculative comment, it is interesting to note some analogy between the source of fronts discussed here and the sources of traveling wave solutions found in non-relaxational flows such as the CGLE[11].

ACKNOWLEDGMENTS: RM and EHG acknowledge financial support from DGYCIT (Spain) Project PB92-0046. R.M. also acknowledges partial support from the Programa de Desarrollo de las Ciencias Básicas (PEDECIBA, Uruguay), the Consejo Nacional de Investigaciones Científicas Y Técnicas (CONICYT, Uruguay) and the Programa de Cooperación con Iberoamérica (ICI, Spain)

**References.**

1- A. Amengual, E. Hernández-García, and M. San Miguel, Phys. Rev. E **47**, 4151 (1993).
2- M. San Miguel, A. Amengual and E. Hernández-García, Phase Transitions **48**, 65 (1994).
3- R. Montagne, A. Amengual, E. Hernández-Gracía and M. San Miguel, Phys. Rev. E **50**, 377 (1994).
4- M. San Miguel and F. Sagués, in *Patterns, defects and materials instabilities*, edited by D. Walgraef and N. Ghoniem (Kluwer, Dordrecht, 1990), and references therein.
5- P. G. de Gennes and J. Prost, *The Physics of Liquid Crystals* (Clarendon, Oxford, 1993).
6- W. van Saarloos, Phys. Rev. A **37**, 211 (1988).
7- P. C. Hohenberg and B. I. Halperin, Rev. Mod. Phys. **49**, 535 (1978).
8- R. Graham, in *Instabilities and Nonequilibrium Structures*, edited by E. Tirapegui and D. Villarroel (Reidel, Dordrecht, 1987), p. 271.

9- R. Graham, in *Theory of continous Fokker-Plank systems*, Vol. 1 of *Noise in nonlinear dynamical systems*, edited by F. Moss and P. V. E. M. Clintock (Cambridge University, Cambridge, 1989), p. 225.
10- R. Graham and T. Tel, in *Instabilities and Nonequilibrium Structures III*, edited by E. Tirapegui and W. Zeller (Reidel, Dordrecht, 1991), p. 125.
11- W. van Saarloos and P. Hohenberg, Physica D **56**, 303 (1992).
12- O. Descalzi and R. Graham, Phys. Lett. A **170**, 84 (1992).
13- O. Descalzi and R. Graham, Z. Phys. B **93**, 509 (1994).
14- B. L. Winkler, H. Richter, I. Rehberg, W. Zimmermann, L. Kramer, A. Buka, Phys. Rev. A **43**, 1940 (1991).
15- M. San Miguel and F. Sagués, Phys. Rev. A **36**, 1883 (1987).
16- F. Sagués and M. San Miguel, Phys. Rev. A **39**, 6567 (1989).
17- A.N. Kolmogorov, I.G. Petrovskii, and N.S. Piskunov, Bull. Univ. Moscou, Ser. Int., Sec. A **1**, 1 (1937), translated in *Dynamics of curved fronts*. P. Pelcé ed. (Academic, San Diego, 1988).
18- W. van Saarloos, Phys. Rev. A **39**, 6367 (1989).
19- G. Dee and J. Langer, Phys. Rev. Lett. **50**, 383 (1983).
20- E. Ben-Jacob, H. Brand, G Dee, L. Kramer and J.S. Langer, Physica D **14**, 348 (1985).
21- W. van Saarloos, Phys. Rev. Lett. **58**, 2571 (1987).
22- J.A. Powel, A.C. Newell and C. K.R.T Jones, Phys. Rev. A **44**, 3636 (1991).
23- S. F. Srajer, G. and R. B. Meyer, Phys. Rev. A **39**, 4828 (1989).